\documentclass[twocolumn,prl]{revtex4}
\usepackage{amsmath}
\usepackage{amstext}
\usepackage{latexsym}
\usepackage{psfig}
\usepackage{graphicx}
\usepackage{amsfonts}
\newcommand{\ket}[1]{|{#1}\rangle}
\newcommand{\bra}[1]{\langle{#1}|}

\begin{document}

\title{Quantum and classical fidelities for Gaussian states}

\author{Hyunseok Jeong}

\author{Timothy C. Ralph}

\address{Department of Physics, University of Queensland, St Lucia, Qld 4072, Australia}

\author{Warwick P. Bowen}

\address{The Jack Dodd 
Centre for Photonics and Ultra Cold Atoms,
Physics Department, University of Otago, Dunedin, New Zealand
}

\begin{abstract}
We examine the physical significance of 
fidelity as a measure of similarity for Gaussian states,
by drawing a comparison with its classical counterpart. 
We find that the relationship 
between these classical and quantum fidelities is not 
straightforward, and in general does not seem to provide
insight into the physical 
significance of quantum fidelity. 
To avoid this ambiguity we propose that
the efficacy of quantum information protocols be 
characterized by determining their transfer function and then 
calculating the fidelity achievable for a hypothetical {\it pure} 
reference input state.
\end{abstract}


\maketitle

\section{Introduction}

Quantification of the similarity (or distinguishability) of
quantum states is a crucial issue in quantum information theory
\cite{nielsen}. {\it Quantum fidelity} \cite{jozsa} - previously
known as Uhlmann's transition probability \cite{uhlmann} - is
probably the most well known such quantification technique, and is
an important tool for assessing the efficacy of quantum
information transfer \cite{braunstein}. Critical to any technique
used to characterize similarity, is a robust understanding of its
physical significance. To date, although there have been efforts
to impose an operational interpretation on quantum fidelity for
mixed states \cite{dodd} and to compare it to alternative distance
measures \cite{lee,gilchrist}, a strong and general physical
significance is yet to be found. When one of the states involved
is pure it is well known that quantum fidelity is equal to the
transition probability from one state to the other. Furthermore,
Uhlmann's theorem allows the quantum fidelity between two
arbitrary states to be translated to a fidelity between higher
dimensional pure states, which can then be interpreted as a
transition probability between those higher dimensional states
\cite{uhlmann}. However, the strength of the link between these
hypothetical higher dimensional states and the actual states under
investigation is not obvious. In this paper we seek to establish
the physical significance of quantum fidelity for a particular
class of states, those of Gaussian nature.

Gaussian states are extremely useful tools in many quantum optics
experiments. For example, several quantum communication
experiments have now been performed using only Gaussian states
\cite{teleport1,teleport2,teleport3,teleport4,lance}. Typically the formulas
used to calculate the quantum fidelity achieved by these
experiments assume that the input states are pure coherent states
\cite{teleport1,teleport2,teleport3,teleport4,lance}. There have been many
studies of quantum fidelity as a success criteria for quantum
teleportation of coherent states \cite{braunstein,tel-fid,tel-fid2,caves},
and its value in this regime is well understood. However, the
unknown quantum states supplied by `Victor' in real experiments
are not perfectly pure, and typically have some small but
non-negligible level of mixedness. It is important to understand
both the effect of this mixedness on the quantum fidelity achieved
by experiments, and the significance of the resulting fidelity
results, which turn out to be markedly different from those
expected for a coherent state even for extremely small levels of
mixedness. Hence the motivation for this paper.

Quantum fidelity is a direct extension of the fidelity between a
pair of classical probability distributions, termed here {\it
classical fidelity}, which is used in statistics to characterize
their similarity. For Gaussian states, in particular, this
relationship is interesting, since the Wigner function describing
such states is non-negative and can be thought of, to some degree,
as a classical probability distribution, which we shall discuss
more rigorously later in this paper. It might then be
expected that the quantum and classical fidelities would coincide
for Gaussian states, and thus a robust physical significance could
be established for quantum fidelity. The results reported here
show, however, that this is not the case.

In this paper, we compare and contrast quantum and classical
fidelities as quantum-classical counterparts.
The explicit forms of the quantum fidelities between two Gaussian states
are obtained for various cases.
We also point out that the classical fidelity between two Gaussian states
inferred by the complementary measurements exactly corresponds to
the overlap between their {\it Wigner functions}.
We then show that although the quantum and classical fidelities do coincide in
the classical limit, i.e., in the limit of the extreme mixedness,
no simple relationship could be established for
non-maximally mixed Gaussian states. The mixedness, squeezing, and
separation of the Gaussian states involved each effect the
discrepancy between classical and quantum fidelity in entirely
different manners.

The unclear physical significance of quantum fidelity between
mixed states raises questions about it's usefulness as a measure
of the efficacy of quantum information protocols. We propose a new
characterization method to avoid this issue. In this method, the
transfer function of the quantum information protocol is
determined, and used to characterize the quantum fidelity
achievable for an arbitrary {\it pure} input state. This method
has two advantages, 1) it provides a standard benchmark through
which to compare different experiments; and 2) the input state can
be chosen to ensure that one of the states used to determine the
quantum fidelity is pure, yielding a physically significant
fidelity, which represents the transition probability from one
state to the other.

\section{Motivation}
\label{motiv}

As we have noted in the introduction the unknown quantum states
used in real continuous variable quantum teleportation experiments
\cite{teleport1,teleport2,teleport3,teleport4} are not
exactly pure states but have some small level of mixedness.
In most experiments, the input states have been assumed 
pure in assessing the efficacy of the quantum information protocols
without justification (we shall further clarify
and discuss this point in Sec.~\ref{QIprot}).
A natural question here is: ``How sensitive is fidelity to
small levels of mixedness?'' It is generally assumed that
the small level of mixedness involved will not significantly change the
fidelity between the input and output states.
However, as we will see here, in general
this turns out not to be the case.

Let us consider the ``no-entanglement'' fidelity limit for unity
gain quantum teleportation of coherent states. This is generally
accepted to be given by a protocol in which Alice makes an ideal
heterodyne measurement of the unknown state, thus obtaining an
estimate of its coherent amplitude, $\alpha$; passes the
measurement result to Bob who then displaces a local vacuum mode
by this estimated value \cite{teleport1}. Bob's state then has the
same average coherent amplitude as the unknown state but its
variance is increased by two units of vacuum noise. If the input
state is a pure coherent state then the no-entanglement fidelity
is $F_{ne}=0.5$. Both the fidelity of quantum teleportation
and the ``no-entanglement'' fidelity limit depend on the class of
input states used.   The degradation of fidelity in
``no-entanglement'' quantum teleportation is a result of noise
introduced to the output state during the measurement and
reconstruction processes.  In the case when the input states are
pure, this noise causes a significant change to the Wigner
function describing the output state, with the result of poor
overlap, fidelity, between the input and output states.  However,
as the input states become more and more thermal, and hence their
breadths become larger and larger, the noise plays a smaller and
smaller role in the overlap between input and output states.  In
the limit that the magnitude of noise introduced is insignificant
compared to the breadth of the input state, the overlap
between the input and output states, and hence
the ``no-entanglement'' fidelity limit, approaches unity.
Our question in this section is how rapid this transition from 0.5
to 1 is as the input states become thermalized.

\begin{figure}
\centerline{\scalebox{.7}{\includegraphics{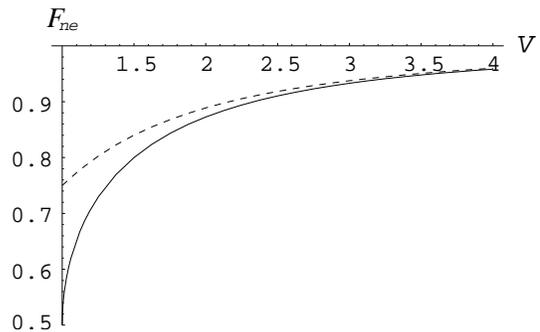}}} 
\caption{The fidelity limit $F_{ne}$ for quantum teleportation
without entanglement against the variance $V$
of an isotropically mixed state input (solid line)
and the corresponding classical fidelity (dashed line).}
\label{fig0}
\end{figure}

A general expression for the fidelity in this situation can easily
be obtained by invoking Uhlmann's theorem \cite{uhlmann}. It
states that if $\rho_{1}={\rm
Tr}_{a}\{\ket{\psi_{1}}\bra{\psi_{1}}\}$ and $\rho_{2}={\rm
Tr}_{a}\{\ket{\psi_{2}}\bra{\psi_{2}}\}$, where $\ket{\psi_{1}}$
and $\ket{\psi_{2}}$ are pure two mode states and the partial
traces are only taken over one of the modes,
 then
the fidelity of $\rho_{1}$ with respect to $\rho_{2}$ is given by
\begin{equation}
     F_{\rho_{1},\rho_{2}}
     =^{max}_{\psi_{1,2}}|\bra{\psi_{2}}\ket{\psi_{1}}|^{2}
\end{equation}
where the maximization is over all pure states which have the
required reduced density operators.   In our case, the reduced
density operators must be those describing the input and output
states, both of which are isotropically mixed coherent states.
Here we limit our analysis to unity gain teleportation, in which
case the average coherent amplitudes of the input and output
states are equal.   We can then choose $\alpha_{1}=\alpha_{2}=0$
without loss of generality, and the input and output states then
become thermal.   In general, 
to calculate the fidelity between input and output 
states we must
then maximize the fidelity over all higher dimensional states
which have reduced density operators describing the required
thermal states. However, it is well known that the partial trace
over an EPR (two-mode squeezed) state, given by
\begin{equation}
     \ket{\phi_{i}} = {{1}\over{G_{i}}} \sum_{n}{
     \left({{(G_{i}-1)}\over{G_{i}}}\right)^{n/2} \ket{n}_{a}\ket{n}_{b}},
\end{equation}
where $G_{i} \ge 1$ is the strength of the squeezing, results in a
thermal state \cite{Walls}. From symmetry it is clear that such a
choice will maximize the fidelity as required. Thus the fidelity
between two thermal states is given by
\begin{eqnarray}
     F & = & |\bra{\phi_{1}}\ket{\phi_{2}}|^{2} \nonumber\\
     & = & {{1}\over{G_{1} G_{2}}} \left(\sum_{n}
     \left({{(G_{1}-1)(G_{2}-1)}
     \over{G_{1}G_{2}}}\right)^{n/2}\right)^{2}\nonumber\\
     & = &
  \left({{1}\over{\sqrt{G_{1}G_{2}}-
     \sqrt{(G_{1}-1)(G_{2}-1)}}}\right)^{2}\nonumber\\
     & = &
   \left({{2}\over{\sqrt{(V_{1}+1)(V_{2}+1)}-
   \sqrt{(V_{1}-1)(V_{2}-1)}}}\right)^{2}
   \label{eq:mix}
\end{eqnarray}
where $V_{i} = 2 G_{i} - 1$ is the variance of the single mode
thermal state obtained by the partial trace of the corresponding
EPR state.

We are now in a position to calculate the no-entanglement fidelity
limit for teleportation of an isotropically mixed coherent state.
Using Eq.~(\ref{eq:mix}) and the fact that the ideal heterodyne
protocol discussed above adds two units of vacuum noise to the
output we obtain
\begin{equation}
     F_{ne} = \left({{2}\over{\sqrt{(V+1)(V+3)}-
   \sqrt{(V-1)(V+1)}}}\right)^{2}.
\end{equation}
This expression is graphed as a function of the input variance $V
\ge 1$ in Fig.~\ref{fig0}. Notice that we recover $F_{ne} = 0.5$
at $V=1$, i.e. a pure coherent state, but that the fidelity is
very sensitive to small amounts of mixedness. For example for 2\%
mixedness, i.e. $V = 1.02$, a typical level of experimental purity,
we have $F_{ne} = 0.57$, a 14\% increase in the classical limit.
We see, therefore, that to ensure accurate 
fidelity results it is critical that the analysis of quantum 
teleportation experiments takes into account the mixedness of the 
input state.  Just as important as achieving an accurate fidelity 
result, is to understand the actual significance of the result.  One 
way of doing this is to consider the analogous classical system.  The 
comparison of quantum and classical fidelities is the topic of 
Section III, and the classical fidelity  (see Eq. (11)) between 
probability distibutions equivalent to the Wigner functions of the 
teleporter intput and output states is also plotted in Fig.~\ref{fig0}.
Notice that for low levels of mixedness the classical and quantum fidelities 
are in stark constrast, but has the mixedness increases they 
asymptote to the same value.

\section{
Comparison between quantum and classical fidelities
for Gaussian distributions}
\label{comparison}

\subsection{General expressions for fidelities of Gaussian states}

\label{qcf}

Classical fidelity $F_c$ and quantum fidelity $F_q$ are defined as
\cite{nielsen}
\begin{eqnarray}
F_c(P_1,P_2)&=&\Big(\int d^2\alpha
\sqrt{P_1(\alpha) P_2(\alpha)}\Big)^2, \label{eqP}\\
 F_q(\rho_1,\rho_2)&=&\Big\{{\rm Tr}\Big[
 \sqrt{\sqrt{\rho_1}\rho_2\sqrt{\rho_1}}\Big]\Big\}^2,
\label{eqR}
 \end{eqnarray}
 where $P_1$ and $P_2$ are probability distributions,
 and $\rho_1$ and $\rho_2$ are density matrices. These density matrices can be equivalently 
represented as quasi-probability distributions, such as the Wigner function.
The Wigner function of a general Gaussian state is
\begin{eqnarray}
W(\alpha)=\frac{2}{\pi\sqrt{V^+V^-}}
\exp\Big[-\frac{2}{V^+}(\alpha_r\cos\phi+\alpha_i\sin\phi-\delta_r)^2\nonumber\\
-\frac{2}{V^-}(\alpha_i\cos\phi-\alpha_r\sin\phi-\delta_i)^2\Big].
~~~~~~~~~~
\label{wigner}
\end{eqnarray}
where $V^+=[\Delta X(\phi)]^2$, $V^-=[\Delta P(\phi)]^2$,
$\hat X(\phi)=(e^{-i\phi}{\hat a}+e^{i\phi}{\hat a}^\dagger)/2$ and
$\hat P(\phi)=-i(e^{-i\phi}{\hat a}-e^{i\phi}{\hat a}^\dagger)/2$.
Note that the variances $V^\pm$ are directly measurable values
in experiments.
Eq.~(\ref{wigner}) becomes a coherent state of amplitude
 $\delta~(=\delta_r+i\delta_i)$ when $V^+=V^-=1$.
Here the breadth of the distribution is quantified by the
product $V^+V^-$.  For quantum states this corresponds directly to
the mixedness of the state, where $V^+V^-=1$ for a pure state and
$V^+V^- \rightarrow \infty$ as the mixedness increases. The level
of squeezing of a Gaussian state is determined by the squeezing
parameter $r$
\begin{equation}
r=\frac{1}{4}e^{i\phi}\ln[\frac{V^-}{V^+}].
\label{R}
\end{equation}
 We will compare two Gaussian distributions
 labelled by the subscripts 1 and 2. These Gaussian distributions
correspond to Wigner functions for quantum fidelity and
 to probability distributions for classical fidelity.

Suppose an ensemble of a Gaussian quantum state $\rho$.
One can measure $\hat X(\tilde\phi)$ many times while varying angle $\tilde\phi$ to find the 
squeezed angle $\phi$. Furthermore, the accumulated measurement results for $\hat X(\phi)$ and 
$\hat P(\phi)$ will result in the Gaussian probability distributions $P(\alpha_r^\prime)$ and 
$P(\alpha_i^\prime)$, where
\begin{eqnarray}
\alpha_r^\prime&=&\alpha_r\cos\phi+\alpha_i\sin\phi,\label{new1}\\
\alpha_i^\prime&=&\alpha_i\cos\phi-\alpha_r\sin\phi,\label{new2}
\end{eqnarray}
with variances $V^+$ and $V^-$.
A classical probability distribution $P(\alpha^\prime)$ for $\alpha_r$ and $\alpha_i$ constructed 
from $P(\alpha_r^\prime)$ and $P(\alpha_i^\prime)$ will be identical to the Wigner function 
$W(\alpha)$ in Eq.~(\ref{wigner}), which can also be represented by $W(\alpha^\prime)$ with the 
rotated variable $\alpha^\prime$.
 {\it In other words, if one infers the two-dimensional classical probability distribution from the 
repetitive complementary
measurements on the ensemble of the Gaussian quantum state $\rho$, it will be exactly the same as 
the Wigner function of state $\rho$.}
{\it Therefore, the Wigner functions should be used to calculate the classical fidelity for 
Gaussian states.}
{\it It should be noted that other quasi-probability distributions such as the $P$-function or 
$Q$-function cannot be used for this purpose.}
Of course, this approach cannot be generally applied to non-Gaussian states which may have 
negativity in the Wigner functions. A similar treatment can be found in a recent work~\cite{cf}, 
where the author showed that a good estimate of the fidelity of a quantum process can be obtained 
by measuring the outputs for only two complementary sets of input states.
 In short, the Wigner function of a Gaussian state 
is a good analogy of the classical probability distribution
{\it in the phase space}, and Eq.~(\ref{eqP}) with the Winger functions
can be a reasonable measure of the classical similarity of two Gaussian states.
On the other hand, it is nontrivial to represent quantum fidelity 
between mixed states in terms of their Wigner functions.
However, when one of the states is pure,
quantum fidelity in Eq.~(\ref{eqR})
can be expressed in terms
of the Wigner functions as
$F_q(\rho_1,\rho_2) =\pi\int d^2\alpha W_1(\alpha)W_2(\alpha)$
\cite{f-1,f-2,f-3}.
It is interesting to note that this formula is
obviously different from classical fidelity in Eq.~(\ref{eqP}).

\begin{figure}
\centerline{\scalebox{.65}{\includegraphics{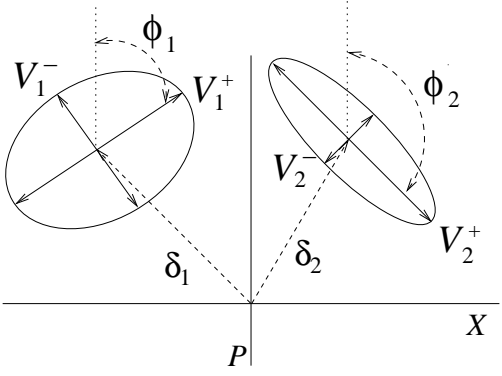}}}
\caption{Schematic of two arbitrary Gaussian distributions $P_1$ and $P_2$.}
\label{fig1}
\end{figure}

Let us first consider the case where $\delta_{1}=\delta_{2}$,
{\it i.e.}, the two Gaussian distributions have the same center.
If we interpret the Wigner function (\ref{wigner}) as a probability
distribution, the classical fidelity between the
Gaussian distributions $P_1(\alpha)$ and $P_2(\alpha)$
is straightforwardly obtained by Eq.~(\ref{eqP}) as
\begin{equation}
\begin{aligned}
&F_c=4\sqrt{V_1^+V_1^-V_2^+V_2^-}
\Big\{\cos^2 \varphi (V_1^+ + V_2^+)(V_1^- + V_2^-)\\
&~~~~~~~~~~~~~~ + \sin^2 \varphi
(V_1^+ + V_2^-)(V_1^- + V_2^+)\Big\}^{-1}
\label{fcp}
\end{aligned}
\end{equation}
where $\varphi=\phi_2-\phi_1$ is the angle between the Gaussian distributions (see 
Fig.~\ref{fig1}).
The quantum fidelity between Gaussian states have been studied by
several authors in previous work and
some useful analytical expressions have been found \cite{twamley,wang-f,scu,MMC}.
It is possible to
transform their formulas into more experimentally useful forms in
terms of $V^{\pm}_{1,2}$ and $\varphi$. The quantum fidelity
is found to be (see Appendix)
\begin{equation}
F_q=\frac{2}{\sqrt{4\sqrt{V_1^+  V_2^+  V_1^-  V_2^-}/F_c +  K} - \sqrt{K}},\\
\label{fcq}
\end{equation}
where $K = (V_1^+  V_1^- -1)(V_2^+  V_2^- -1)$. For most of
the cases considered in this paper the angle $\varphi$ is zero, in
this case the classical and quantum fidelities are
\begin{eqnarray}
&&F_c(\varphi=0)=\frac{4\sqrt{V_1^+V_1^-V_2^+V_2^-}}
{(V_1^++V_2^+)(V_1^-+V_2^-)},\\
&&F_q(\varphi=0)=2 \Big\{\sqrt{(V_1^+V_2^-+1)(V_1^-V_2^++1)}
-\sqrt{K}\Big\}^{-1}.\nonumber\\
\end{eqnarray}
Even simpler formulas result when the amplitude and phase
quadratures are symmetric ($V_1^+=V_1^-=V_1$ and
$V_2^+=V_2^-=V_2$), in that case
\begin{eqnarray}
\label{cfforc1}
&&F_c(V_1,V_2)=\frac{4V_1V_2}{(V_1+V_2)^2},\\
\label{jts}
&&F_q(V_1,V_2)=2\Big\{V_1V_2+1-\sqrt{(V_1^2-1)(V_2^2-1)}
\Big\}^{-1},~~~~~~ 
\end{eqnarray}
where Eq.~(\ref{jts}) and Eq.~(\ref{eq:mix}) are identical.

In general, the two states can be separated by 
some distance $x=x_r + i x_i = \delta_2 - \delta_1$ in phase 
space.  
This separation 
can be shown to have the following straightforward effect (see Appendix):
\begin{eqnarray}
\begin{aligned}
&&F_q(x)=F_q(\varphi=0){\cal D}(x),~~~~~~~~~~~~~~~~~
\label{fqd}\\
&&{\cal D}(x)=\exp[-\frac{2x_r^2}{V_1^++V_2^+}-\frac{2x_1^2}{V_1^-+V_2^-}].
\end{aligned}
\end{eqnarray}
This dependence on distance
turns out to be exactly the same as one obtains for classical
fidelity
\begin{equation}
F_c(x)= F_c(\varphi=0){\cal D}(x),
\end{equation}
and we therefore consider 
the separation in phase space no further here

\subsection{$\rho_{1}$ or $\rho_{2}$ pure}

\begin{figure}
\centerline{\scalebox{.63}{\includegraphics{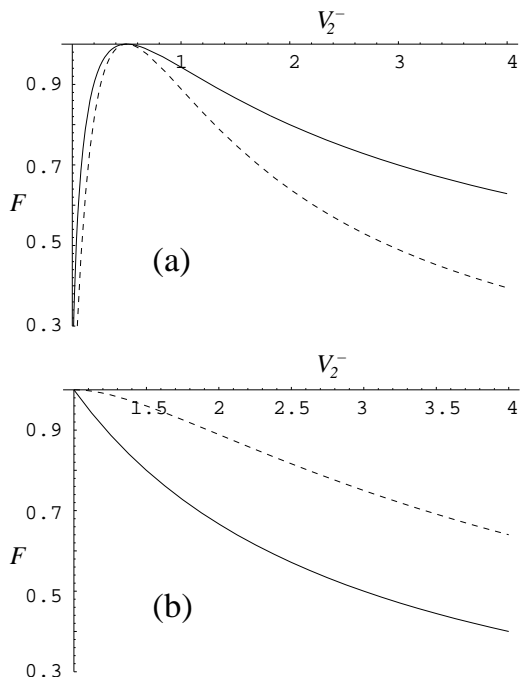}}}
\caption{Quantum (solid line) and classical (dashed line)
fidelities, $F$, between two Gaussian distributions.
 (a) Both distributions pure, $V_1^+=2$, $V_1^-=1/2$, $V_2^+ V_2^-=1$ and $\varphi=0$.
(b) One distribution pure, and the other mixed but with the same
squeezing parameter $V_1^+=2$, $V_1^-=1/2$, $V_2^+/V_2^-=4$ and
$\varphi=0$.}
\label{fig2}
\end{figure}

Let us first compare $F_c$ and $F_q$ from
Eqs.~(\ref{fcp}) and (\ref{fcq}) when one of the states is pure.
In this case, a simple relationship can be drawn between the
quantum and classical fidelities
\begin{equation}
F_q^2=\frac{F_c}{\sqrt{V_2^+ V_2^-}}. \label{FidelitySimple}
\end{equation}
Here, as in all following cases, when comparing classical and
quantum fidelities the properties of distribution 1 are fixed while
those of distribution 2 are varied. The quantum and classical
fidelities between two distributions with $V_1^+=2$, $V_1^-=1/2$,
$V_2^+ V_2^-=1$ and $\varphi=0$ are compared in
Fig.~\ref{fig2}(a). For quantum fidelity, this condition
corresponds to two pure quantum states, one of which has a varying
degree of squeezing while the other has a fixed squeezing
parameter of $r=-0.347$. We see that classical fidelity degrades
faster than quantum fidelity.   This result can be obtained
straightforwardly from Eq.~(\ref{FidelitySimple}).  We see that
when $V_2^+ V_2^-=1$, $F_c=F_q^2$; remembering the bound on
fidelity $0 \le \{F_c,F_q\} \le 1$ it is clear that $F_c \le
F_q$.

Let us now consider the quantum and classical fidelities
between a pure squeezed state, and a mixed state with the same
squeezing parameter.  Results for the parameters $V_1^+=2$,
$V_1^-=1/2$,
 $V_2^+/V_2^-=4$, $r=-0.347$, and $\varphi=0$ are shown in
 Fig.~\ref{fig2}(b). We see that the quantum fidelity degrades
 faster than the classical fidelity as the difference in the
 mixedness of the two states increases.   This trend directly
 contrasts that obtained in Fig.~\ref{fig2}(a) as the discrepancy in the squeezing parameters of 
the
 two states increases.

Fig.~\ref{fig2}(b) seems to reflect the difference
between quantum and classical distributions with respect to
mixedness.  Since a classical ensemble consists simply of a
mixture of classical states each weighted by the distribution
function, it is essentially entirely mixed regardless of the
breadth of the distribution.  A quantum state however, is
perfectly pure if the breadth is unity ($V^+V^-=1$), and the
mixedness increases as $V^+V^- \rightarrow \infty$.   So,
increasing the breadth of the distribution has a secondary effect
for quantum states which is not present for classical
distributions. It is reasonable, then, that differences in breadth
cause a greater reduction in the similarity (and hence fidelity)
of quantum states than that of their classical counterparts under
the same conditions.

\subsection{$\rho_{1}$ and $\rho_{2}$ mixed}

As we have explained in Sec.~\ref{motiv}, it may not be acceptable
to simply assume that the input state is pure in real
teleportation experiments, since the quantum fidelity can be
extremely sensitive to even small amounts of mixedness. For
example, the quantum fidelity between a coherent state of
$V_1=1$ and a thermal state of $V_2=2$ is
$F\approx0.667$, whereas for thermal states with
$V_1=1.05$ and $V_2=2$
($V_1=1.01$ and $V_2=2$) it is
$F\approx0.785$ ($F\approx0.721$).
 If a pure input state was
assumed for the latter cases the fidelity would be only
$F\approx0.667$, a significantly underestimation.

\begin{figure}
\centerline{\scalebox{.63}{\includegraphics{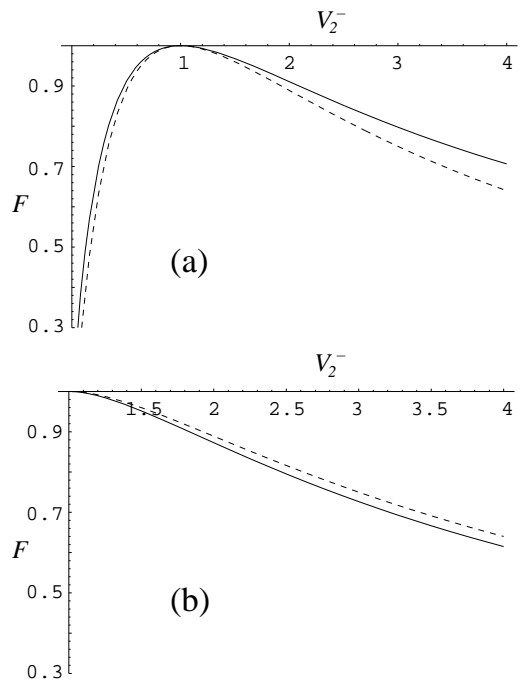}}}
\caption{Quantum (solid line) and classical (dashed line)
fidelities between Gaussian distributions. (a) $V_1^+=4$,
$V_1^-=1$, $V_2^+ V_2^-=4$ and $\varphi=0$. (b) $V_1^+=4$,
$V_1^-=1$, $V_2^+/V_2^-=4$ and $\varphi=0$. The breadths of
the Gaussian distributions are four times larger than those in
Fig.~3.}
\label{fig3}
\end{figure} 

In Fig.~\ref{fig3}, the breadth  of the distributions are
four times larger than those for the previous case
(Fig.~\ref{fig2}), while all the other conditions are same.
The same trends to those in Fig.~\ref{fig2} are observed but the
differences between quantum and classical fidelities are smaller.
In other words, the discrepancy between quantum and classical
fidelity is reduced as the breadth ($V^+ V^-$) of the Gaussian
distributions increases.  The two fidelities become identical as
$V^+_1V^-_1\rightarrow\infty$ and $V^+_2V^-_2\rightarrow\infty$.
Since in this limit the quantum states can be treated as classical
objects we see that the classical limit of quantum fidelity is
classical fidelity as expected.

\begin{figure}
\centerline{\scalebox{.65}{\includegraphics{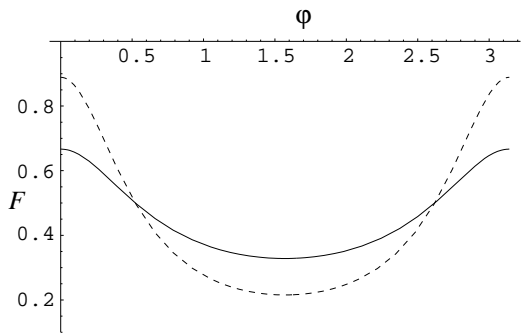}}}
\caption{Quantum (solid line) and classical (dashed line)
fidelities for two Gaussian distributions having the same absolute
squeezing parameter, $V_1^+/V_1^-=V_2^+/V_2^-=16$
($|r|=0.693$), yet different breadths
$V_1^+V_1^-=V_2^+V_2^-/4=1$. The Quantum and classical fidelities
vary differently with the relative angle.}
\label{fig4}
\end{figure}

As a final example, let us consider the effect of rotating
one distribution in phase space. Suppose the two Gaussian
distributions have the same absolute squeezing parameter,
($V_1^+/V_1^-=V_2^+/V_2^-=16$, $|r|=0.693$), but different breadths
($V_1^+V_1^-=V_2^+V_2^-/4=1$),
and are initially aligned ($\varphi=0$).
In this case, as was seen previously, the
difference in breadth causes the classical fidelity to be greater
than the quantum fidelity (similar to Fig.~\ref{fig2}(b)).
However, if one begins to change the relative angle $\varphi$
between the distributions,
the squeezing parameters, $r_1$ and $r_2$,
of the distributions become different.
At a certain point, this difference may become more dominant than
the difference in breadth (similar to Fig.~\ref{fig2}(a)).
Thus the difference between
quantum and classical fidelities gets smaller and eventually
classical fidelity becomes greater as shown in Fig.~\ref{fig4}.

In this Section, we have considered several different manipulations of 
Gaussian states, and shown that quantum and classical fidelities 
respond in qualitatively different manners to these 
manipulations.  
It is clear, therefore, that classical fidelity 
cannot be used, in general, to establish a strong physical 
significance for quantum fidelity. 
As mentioned previously, there is 
a clear physical significance if one of the states involved is 
pure. 
In the next Section, we propose a characterization technique 
that takes advantage of this fact to establish a physical 
significance for the fidelity of general quantum information protocols.

\section{Fidelity for quantum information protocols}
\label{QIprot}

One of the most common applications of quantum fidelity is to measure the efficacy of quantum 
information protocols \cite{nielsen}.   Typically, to characterize such protocols one begins with 
an ensemble of identical known input states.  The protocol is then performed on each input state, 
yielding an ensemble of (hopefully identical) output states.   These states can be fully 
characterized by performing tomographic measurements.   The desired output state 
(the output state that would be achieved if the protocol ran perfectly)
from the protocol is typically well known.
For example, in unity gain quantum teleportation it would simply be the input state. The fidelity 
between this  desired output state and the actual output state can then be directly calculated, and 
is used to judge the efficacy of the protocol.   However, we have seen that fidelity is highly 
sensitive to both squeezing of, and impurity in, the states being compared.   Furthermore, when 
both states are impure, it is difficult to attribute anything more than a weak physical 
significance to fidelity.   Since in any realistic experiment all states involved (except perhaps 
vacuum states) will be at least somewhat impure, the usefulness of fidelity per se as a measure of 
the efficacy of quantum information protocols is questionable.  Furthermore, since the fidelity of 
a quantum information protocol depends strongly on the properties of the input state, care must be 
taken when using fidelity to make comparisons between even very similar experiments.

At this point one would be forgiven for contemplating rejecting fidelity entirely as an efficacy 
measure.   However, it does have one very attractive feature:  when at least one of the states 
involved is pure, it is simply the transition probability between the two states.   In other words, 
if one was to make a projective measurement on the output state, fidelity is the probability that 
it would collapse into the desired output state.
An alternative technique, that we advocate here, is to characterize the transfer function of the 
quantum information protocol.   In general, this will involve characterizing the function that maps 
the Wigner function of the input state to that of the output state. Given particular experimental 
conditions and available input states, this can be achieved to the same particular precision.  The 
fidelity for any arbitrary input state can then be calculated simply by applying the transfer 
function, and then comparing the resulting predicted output state to the desired output state.   
The fidelity for some {\it reference input state}, which can be chosen arbitrarily,  can then be 
calculated. Fidelity calculated in this way has the following advantages: 1) since a pure input 
state can be chosen, the desired output state from the protocol will also often be pure,
\cite{foot1}
the fidelity obtained then has physical significance as the transition probability between the 
desired and predicted output states; and 2) since the same reference input state can be used for 
all experimental implementations of quantum information protocols, this fidelity can be used as a 
benchmark to
compare efficacies.

Let us consider, for example, how this process would work for unity gain continuous variable 
quantum teleportation.   As mentioned above, the desired output  for unity gain continuous variable 
quantum teleportation is the input state.  However, as a result of imperfect entanglement and 
technical noise sources the actual output state will be somewhat degraded.   This effect is 
quantified by the transfer function of the system; defined by the gain of the process which due to 
experimental imperfections will not be exactly unity \cite{teleport2,teleport3}, and the additional 
noise present on the output state over-and-above the noise present on the input state.    In 
general the additional noise is non-Gaussian and tomographic techniques are required for a full 
characterization, however when Gaussian entanglement is used and all other noise sources are 
Gaussian only the variance of the noise is required \cite{teleport2,teleport3,RalphLam}.   This is 
the case for all continuous variable teleportation experiments to date 
\cite{teleport1,teleport2,teleport3,teleport4}.   Once the gain and noise variance have been 
determined an arbitrary reference input state can be
chosen and the corresponding output state calculated.   A sensible
choice of reference input state in this case would be a coherent
state, since the classical fidelity limit for teleportation is
normally quoted for coherent states.   The fidelity between this
reference coherent state and the output state can be directly
calculated, and used to compare different teleportation
experiments.  As we saw in Fig.~\ref{fig0}, if this transfer
function and reference state technique is not used to characterize
teleportation experiments, the fidelities quoted by different
experiments will vary significantly based on variations in the
mixedness of the input states.  This variation has no bearing on
the strength of entanglement used in the experiment, or the
efficacy of the protocol. Note that in most teleportation
experiments to date it has simply been assumed their input states
were pure without justification. For small levels of mixedness,
this is approximately (but not exactly) equivalent to what we suggest here.

\subsection{A specific example: comparison of two unity gain teleportation experiments}

Consider two unity gain teleportation experiments, each performed
in exactly the same manner, and each with identical entanglement
resources. Let us say, for arguments sake, that the entanglement is
generated by interfering two squeezed beams (labelled here with
subscripts $a$ and $b$, respectively) with variances $V_a^+ =
V_b^-=1/V_a^-=1/V_b^+ = 0.5$. Assuming that, apart from non-ideal
entanglement, both experiments are performed perfectly,
\cite{foot2}  the output of each is a Gaussian
state with amplitude and phase quadrature variances given by
\begin{equation}
V^{\pm}_{\rm out} = V^{\pm}_{\rm in} + 1, \label{TeleEq}
\end{equation}
where $V^{\pm}_{\rm in}$ are the amplitude and phase quadrature
variances of the input state \cite{teleport2,teleport3}. Notice, that the
noise introduced to the output state is entirely independent of
the input state. It should therefore be concluded that both
teleportation experiments performed equally well. However, lets
say that the first experiment used a coherent input state ($V^{\pm
(1)}_{\rm in}=1$), whilst the second used a thermal state with
$V^{\pm (2)}_{\rm in}=2$. The output states then have respective
variances of $V^{\pm (1)}_{\rm out} = 2$, and $V^{\pm (2)}_{\rm
out} = 3$. Substituting these values directly into
Eq.~(\ref{eq:mix}) ($V^{\pm (i)}_{\rm in} \rightarrow V_1$,
$V^{\pm (i)}_{\rm out} \rightarrow V_2$) we see that the
experiments yield dramatically different fidelities of
$F^{(1)}=0.67$ and $F^{(2)}=0.95$, and an incorrect conclusion
could be drawn that experiment $(2)$ performed much better than
experiment $(1)$.

If the fidelity is calculated via a transfer function approach,
however, this difference is eliminated. As discussed previously,
the transfer function of a teleportation experiment for which the
entanglement and noise sources are Gaussian can be characterized
simply by the teleportation gain, and the variance of the
introduced noise. Experimenters $(1)$ and $(2)$, therefore, both
determine these parameters from measurements of the coherent
amplitudes and variances of their respective input and output
states. In both cases the gain and noise variances will be equal
to unity as can be seen from Eq.~(\ref{TeleEq}); and
Eq.~(\ref{TeleEq}) then directly defines the transfer function of
the teleportation system. To compare experiments, the
experimenters choose a common reference input state, in this case
a coherent state, and determine from Eq.~(\ref{TeleEq}) that, if
such an input state was used in their system, the output variances
would be $V^{\pm}_{\rm out}=2$. They then arrive at the fidelity
of teleportation for this particular reference input state from
Eq.~(\ref{eq:mix}), which yields a value of $F=0.67$ in both
cases. The experimenters therefore reach the correct conclusion
that their experiments were performed equally well.

\section{Remarks}
\label{remarks}

In this paper we have investigated the quantum fidelity
between Gaussian states.  Investigations of this kind are
important, since all continuous variable quantum information
experiments to date have been performed with such states.  The
input states in these experiments are normally treated as
pure coherent states \cite{teleport1,teleport2,teleport3,teleport4}.
However, small levels of mixedness are typically present.  We find
that even these levels of mixedness significantly alter the quantum
fidelity. Hence, it is typically not appropriate to simply assume
that the input states are pure.  In an attempt to understand
why quantum fidelity is so sensitive to mixedness, and to gather
some understanding of its physical significance between two mixed
states, we consider its classical counterpart, the classical
fidelity between two probability distributions. Since the Wigner
functions of Gaussian states are positive definite, one might
expect the quantum and classical fidelities to be identical or at
least similar. We find, however, that they show radically different
behaviors. Classical fidelity between probability distributions is
degraded more strongly than quantum fidelity between quantum states
as a result of differences in squeezing parameters. On the other hand,
the quantum fidelity degrades faster than the classical fidelity
as the breadth ($\Delta X^2 \Delta P^2$) of the distributions
diverge. The distance between two Gaussian states in phase space
does not cause any discrepancy between the quantum and classical
fidelities.
In the limit of the extreme mixedness for both Gaussian states,
which can be considered the classical limit,
quantum fidelity approaches the classical one.

Although a clear physical significance can be attached to quantum
fidelity when one of the states involved is pure, our results
indicate that when both states are mixed, quantum fidelity loses
this significance.    For this reason, we propose the use of
transfer functions to characterize continuous variable quantum
information protocols.  Once the transfer function of the protocol
is determined, the fidelity that would be achieved between an
arbitrary pure input state, and the output state can be
calculated.   The resulting value has physical significance and
can be used as a benchmark to compare between experiments.

\appendix

\section*{Appendix A: Quantum fidelity for Gaussian states}

\setcounter{equation}{0}
\renewcommand{\theequation}{A{\arabic{equation}}}

The density matrix of a general Gaussian state can be expressed as
\begin{equation}
\rho=Z(\beta) D(x)S(r)\exp[-\frac{\beta}{2}(a a^\dagger+a^\dagger
a) ]S^\dagger(r)D^\dagger(x), \label{GR}
\end{equation}
where $S(r)$ is the squeezing operator, $D(x)$ is the displacement
operator, and
  $Z(\beta)$ is the normalization factor.
Quantum fidelity between two Gaussian states $\rho_1$ and
$\rho_2$, for $x_1=x_2$, is then \cite{twamley}
\begin{equation}
F_q^{(\varphi)}=\frac{2\sinh\frac{\beta_1}{2}
\sinh\frac{\beta_2}{2}}{\sqrt{Y}-1}
\label{f-twam}
\end{equation}
where
\begin{widetext}
\begin{eqnarray}
&&Y=\cos^2\varphi\Big[\cosh^2(r_2-r_1)\cosh^2\frac{(\beta_1+\beta_2)}{2}
-\sinh^2(r_1-r_2)\cosh^2\frac{(\beta_2-\beta_1)}{2}\Big]\nonumber
\\
&&~~~~~~~+\sin\varphi\Big[\cosh^2(r_1+r_2)\cosh^2
\frac{(\beta_1+\beta_2)}{2}-\sinh^2(r_1+r_2)
\cosh^2\frac{(\beta_2-\beta_1)}{2}\Big].
\label{Y}
\end{eqnarray}
\end{widetext}
The variances $V^\pm$ for a Gaussian state of the general form
(\ref{GR}) are
\begin{eqnarray}
&&V^+=\Delta X^2=1+A+B,\\
&&V^-=\Delta P^2=1+A-B,\\
&&A=2[{\bar n}+(2{\bar n}+1)\sinh^2r],\\
&&B=2(2{\bar n}+1)\cosh\phi\sinh r\cosh r,\\
&&{\bar n}={\rm Tr}[\rho {\hat a}^\dagger{\hat
a}]=\frac{1}{e^\beta-1},
\end{eqnarray}
where $\bar n$ corresponds to the average photon number. Then the
squeezing parameter $r$ and inverse temperature $\beta$ can be
expressed in terms of $V^{\pm}_{1,2}$ and $\varphi$ as
\begin{equation}
\beta=\ln[1+\frac{2}{\sqrt{V^+V^--1}}],\\
\label{beta}
\end{equation}
and Eq.~(\ref{R}). Eq.~(\ref{fcq}) is obtained from
Eqs.~(\ref{R}), (\ref{beta}), (\ref{f-twam}) and (\ref{Y}).

Quantum fidelity between two distant Gaussian states $\rho_1$ and
$\rho_2$, for $\varphi=0$, was calculated by Wang {\it et al.}
\cite{wang-f} as
\begin{equation}
F_q^{(x)} =F_q^{(\varphi=0)}{\cal D}, \label{AX}
\end{equation}
where
\begin{eqnarray}
\label{BX}
{\cal D}&=&\exp[\frac{(\epsilon_1+\epsilon_2)}{\Delta}],\\
\Delta&=&\cosh\beta_1\cosh\beta_2+\sinh\beta_1\sinh\beta_2\cosh2(r_1-r_2)-1,
\nonumber\\
\epsilon_1&=&\sinh\beta_1\sinh^2\frac{\beta_2}{2}\Big[(g^2+{g^*}^2)\sinh
2r_1
-2|g|^2\cosh 2r_1\Big],\nonumber\\
\epsilon_2&=&\sinh\beta_2\sinh^2\frac{\beta_1}{2}\Big[(g^2+{g^*}^2)\sinh
2r_2 -2|g|^2\cosh 2r_2\Big].\nonumber
\end{eqnarray}
By substituting $\beta$ and $r$  in Eqs.~(\ref{AX}) and (\ref{BX})
with Eqs.~(\ref{beta}) and (\ref{R}), Eq.~(\ref{fqd}) is obtained.

\section*{Acknowledgments}

This work was supported by the Australian Research Council.
WPB acknowledges financial support from the Center for the Physics
of Information, California Institute of Technology.


\begin{thebibliography}{99}
\bibitem{foot1} There are some exceptions, such as non-unity gain teleportation.
\bibitem{foot2} This simplifying assumption has no bearing on the
conclusions of the analysis.
\end{thebibliography}
\end{document}